\documentclass[12pt]{article}
\usepackage[dvips]{graphicx}
\usepackage[dvips]{graphics}

\newcommand{\newc}{\newcommand}
\newc{\lra}{\leftrightarrow}
\newc{\beq}{\begin{equation}}
\newc{\eeq}{\end{equation}}
\newc{\barr}{\begin{eqnarray}}
\newc{\earr}{\end{eqnarray}}
\title {Dark matter search by exclusive studies of X-rays following WIMPs nuclear interactions }
\author{H.~Ejiri$^1$\footnote{\texttt{e-mail:\,
ejiri\,@\,rcnp.osaka-u.ac.jp}}\,,
Ch.~C.~Moustakidis$^2$\footnote{\texttt{e-mail:\,
moustaki\,@\,auth.gr}}\,, and
J.~D.~Vergados$^3$\footnote{\texttt{e-mail:\,
Vergados\,@\,cc.uoi.gr}}
\\
{\it 1 INT, University of Washington, Seattle, WA 98195, USA;} \\
{\it JASRI-SPring-8, Mikazuki-cho, Hyogo, 679-5198}\\
{\it  $^{2}$Department of Theoretical Physics,} \\
{\it Aristotle University of Thessaloniki, }
{\it 54124 Thessaloniki Greece}\\
{\it $^{3}$University of Ioannina, Ioannina, GR 45110, Greece} }
\date{}
\begin{document}
\maketitle
\begin{abstract}
 It is shown that weakly interacting massive particles (WIMPs),
 which are possible cold dark matter candidates,
can be studied by exclusive measurements of X-rays following WIMPs nuclear
 interactions. Inner-shell atomic electrons are ionized through WIMP-nuclear interaction,
 and then mono-energetic X-rays are emitted when they are
 filled by outer-shell electrons. The number of
inner-shell holes amounts to as large as one per five nuclear
recoils for K-shell and several per recoil for L-shell in the case
of medium heavy target nuclei interacting with 100-300 GeV WIMPs.
Then the K and L X-ray peaks show up in the 5-50 keV region.
Consequently exclusive studies of the X-rays in coincidence with
the nuclear recoils and the ionization electrons are found to
provide excellent opportunities to detect WIMPs such as the
Lightest Super Symmetric Particles (LSP)
\end{abstract}
\
\section{Introduction.}
Exotic dark matter is one of the major components of the dark matter in the universe.
 Experimental studies of the
dark matter \cite{san02,mau02,hal02}, together with the
recent WMAP observations \cite{spe03,teg04}, indicate that
the universe consists of the dark matter with $\Omega _{CDM} \approx 30 \%$, the dark energy
with $\Omega _{\Lambda} \approx 65 \%$ and the baryonic matter with
$\Omega _{b} \approx 5 \%$, with $\Omega _{XY}$
 being the fraction of the mass of the type $XY$.
The dark matter is considered to be mainly of the cold variety
made of Weakly Interaction Massive Particles (WIMPs) such as the
Lightest Super-symmetric Particles (LSP). The nature of WIMPs can
only be unveiled by their direct detection in the laboratory.  It
is, thus, of great interest to directly search for WIMPs from the
view point of both particle physics and cosmology. Such
experimental studies of WIMPs, however, require high sensitivity
detectors and/or novel methods,  since the   WIMP signals are
expected to be  extremely rare and occur at very low energy.

The purpose of the present letter is to show for the first time
that:
\begin{enumerate}
 \item[(i)] inner-shell electrons are well excited through WIMPs
nuclear interactions,
 \item[(ii)] hard X-rays are emitted when
inner-shell holes thus created are filled by outer-shell electrons and
\item[(iii)] exclusive measurements of such X-rays provide new excellent
opportunities for high-sensitivity studies of WIMPs.
\end{enumerate}

Many experimental searches for WIMPs have been made in the last
decade. As a result the DAMA experiment has claimed the detection
of cold dark matter particles with a mass of $\approx  30-100$
GeV, by the observation of a seasonal modulation of the low-energy
spectra \cite{ber98}.
On the other hand subsequent experiments, such as the EDELWEISS
\cite{ben02} and CDMS\cite{ake03} data, almost exclude
 cold dark matter events in the region claimed by DAMA. At least, if the WIMP-nuclear interaction is of
the scalar type. In order to settle such issues several groups are currently planning suitable experiments
using high-sensitivity detectors.

 All searches for  WIMPs have so far  been made
by attempting to observe the recoiling nuclei, following their elastic scattering with the WIMPs. This seemed to be the only possibility, since the average energy of the dark matter constituents is too low to excite the nucleus
with appreciable rates. In fact most experiments have employed targets with a large mass number $A$,
expecting the coherent mode to dominate, since the cross section  is proportional to  $A^2$.  In any event the recoil energy
 spectrum of the elastic scattering is of a continuum shape,
 decreasing rapidly as the energy increases.
It thus exhibits  almost identical behavior with  that of the background. It is thus
very hard to separate and identify the interesting recoil signal  from those due to the background.

In order to identify WIMPs signals, one needs to use detectors
which are almost free of backgrounds or to identify and
  measure additional quantities with signatures, which are characteristic of  WIMPs.
 Thus many groups measure the ratio of ionization to phonon,
the ratio of ionization to scintillation and/or the pulse-shape to
differentiate the electron backgrounds. The seasonal
 modulation of the event rates, due to motion of  the earth motion
 is  another good way to identify the cold dark matter,
provided that origin of the backgrounds
is fully understood and one is convinced that it does not show any seasonal variation. Directional
experiments correlated with the sun's direction of motion are also good signatures.  Especially if they are combined with  measurements of the annual
modulation. Admittedly, however, such experiments are hard \cite{ver04}.

It has been shown back in 1993 by one  of the present authors (H.E) and his
colleagues \cite{eji93} and recently by
 another (J.V.D) and his colleague \cite{ver05} that $\gamma $-rays,
following inelastic scattering off nuclei,
provide a unique way to study WIMPs, in particular spin-coupled dark
matter since the $\gamma $ energy can be   as
large as $10 - 100$ keV and is monochromatic. Furthermore, unlike the nuclear recoil
measurements, no quenching effect appears in the case of the
$\gamma $ detection. For these reasons it appears quite feasible  to measure such $\gamma $ rays.

In the previous paper \cite{ver05a,mou05}, we have shown that the
 measurement of ionized electrons via WIMPs
nucleus interactions can be a good and realistic way for the direct detection
 of the LSP.

The present proposal, i.e.  X-ray measurements for the
identification of the dark matter constituents, is based on the
above mentioned previous studies of the $\gamma $-ray and the
ionization electron detection. More specifically in the present
letter we suggest that the X-rays, following the elastic
scattering of WIMPs off medium heavy nuclei, have several
 unique features making them suitable for WIMP studies.  We show, in particular, that the production rate of
 K and L X-rays is large  and, accordingly, the
 direct detection of WIMPs by exclusive studies of the X-rays is realistic.\\
The study  of the X-rays  following the scattering of WIMPs
off nuclei offers some unique advantages such  as:
\begin{enumerate}
\item The K and L X-rays show discrete peaks in the 5-50 keV energy region
for medium - heavy nuclei, in contrast to the nuclear
 recoil detection, which exhibits a continuous spectrum falling rapidly with  energy
  beyond a few keV. The observed energy spectrum
of the nuclear recoil is further shifted to the lower energy
region in a solid detector due to the quenching effect. The X-rays
are thus free from detector threshold effects, since the threshold
is below a few keV in most detectors. \item The nuclear recoil is
followed by hard X-rays. Then exclusive measurement of the nuclear
recoil in coincidence with such hard X-rays reduce most BG signals
by orders of magnitudes. Thus it is realistic with the exclusive
measurement to study WIMPs in the region of around 10$^{-7}$ pb
proton cross section. Detectors for such high-sensitivity
experiments are going to be discussed later.
\item The X-ray production
rate relative to the standard nuclear recoil
 rate is determined by kinematical conditions. It
thus depends on the dark matter particle mass, but it is
independent of any other properties of WIMPs or the nuclear
structure. This is not the case of the $\gamma $-rays arising from
inelastic scattering, since the inelastic cross sections depend on
the nuclear structure and are not well known.
\item X-ray detection
will aid the standard experiments in two ways:
\begin{itemize}
\item by inclusive measurements of the energy sum of the nuclear
recoil, the ionization electrons and the X-rays,
\item by exclusive studies of the X-rays in coincidence with the
 nuclear recoil and/or the ionization electrons. Experimentally,
 identification of the X-ray, the ionization electron
 and the nuclear recoil are feasible by spatial and time correlation
 analyses of their energy deposits.
 \end{itemize}
\end{enumerate}

The X-ray production rate is simply evaluated as in case of the
ionization electron rate discussed in the previous papers
\cite{ver05a,mou05}. Using the same notations there, the ratio of
the $n\ell $-shell electron ionization rate to the nuclear recoil
rate, normalized to one electron per atom, is given as
\cite{mou05}
 \begin{equation}
 \frac {\sigma _{n\ell }}{\sigma _r}=  p_{n\ell } \int ~| \phi _{n\ell }
 (\sqrt {2m_e T)})~|^2 ~f(T,\epsilon_{n\ell }) ~ m_e \sqrt{(2m_eT)}~ dT,
 \label{ionel}
 \end{equation}
 with
 \begin{equation}
 f(T,\epsilon_{n\ell }) = \frac {\int N v^2 e^{-v^2/v^2_0}sinh(2v/v_0)dv}{\int D v^2 e^{-v^2/v^2_0}sinh(2v/v_0)dv}
 \label{intnumden}
 \end{equation}
 where $p_{n\ell}$ is the probability of one electron in the $n\ell $ shell.

Since a real atom has $Z$ electrons per atom, the cross-section
ratio to be observed is obtained by multiplying the ratio given in
Eq. (\ref{ionel}) by the factor $Z$.
 In the last equation, both the numerator  and the denominator
 (nuclear recoil) have been  convoluted with the WIMP velocity distribution, which
 is taken to be Gaussian with respect to the galactic center \cite{mou05}.

 The kinematical ranges have been discussed in our
 earlier work \cite{mou05}. For the reader's convenience we only mention here that the upper  limits of
  integration  in both the numerator and the denominator of Eq. (\ref{intnumden})
   correspond to the maximum LSP velocity, which is given by the escape velocity,
    $\upsilon_{esc}=2.84\upsilon_0$ with $\upsilon_0=220~km/s$ the sun's rotational
    velocity. The lower limit in the denominator is set to zero, a conservative estimate
    assuming that the nuclear recoil detection can go down to
 zero energy threshold. The lower limit in the numerator is given by:
 \beq
\upsilon_{min}=\sqrt{ \frac{ 2 \left( T-\epsilon_{n \ell}
\right)}{\mu_r} } \label{vmin}. \eeq
 In other words the whole
expression is a function of $T$ and $\epsilon_{n \ell}$. The range
of integration in Eq. (\ref{ionel}) over the  electron energy is
between  $T=0$ and
$T=\frac{1}{2}m_{\chi}\upsilon^2_{esc}+\epsilon_{n \ell}$. The
precise value of the maximum energy is not very relevant, since
the rate peaks at much lower electron energies. If one attempts to detect electrons one
encounters a threshold energy \cite{mou05},  $E_{th}$, but in the experiment proposed here
 X-rays will be detected, not electrons. For completeness, however, we mention that the total rates for electron detection  are sensitive to the value of $E_{th}$ \cite{mou05}, but the rates associated with the inner
 shell electrons are not.
 This is true in particular for the $1s$
 rates, regardless of the WIMP mass. For a WIMP mass of $100$ GeV the $1s$
 rates change less than  $3\%$ in going from $E_{th}=0$ to $E_{th}=0.2$ keV.

 The inner-shell production rates have been  evaluated  in the case of
 the $^{131}$Xe isotope, as a typical isotope of medium heavy nuclei.
It is noted that inner-shell excitation rates depend on the atomic
number, as it has previously been shown \cite{mou05}, but not on
the mass number. In fact natural Xe isotopes will be used for
experiments with Xe. The ratios of the inner-shell ionization to
the nuclear recoils, together with the binding energies, are shown
for light, medium and heavy WIMPs with the mass of 30, 100 and 300
GeV in Table \ref{t:1}.

 \vspace{0.5 cm}
 \begin{table}[h]
\begin{center}
\caption{The binding energies and the inner-shell ionization
ratios in WIMP nuclear interactions for $^{131}$Xe. The
inner-shell ionization rates, normalized to one electron per atom,
relative  to the nuclear recoil rates are given in the 3-5
columns, and the inner-shell cross-sections relative to the
nuclear recoil cross-sections are in the 6-8 columns.
}
 \label{t:1}
\vspace{0.5cm}
\begin{tabular}{|cccccccc|}
\hline $n \ell$   & $-\epsilon _{n\ell }$ keV & $[\frac{\sigma
_{n\ell }}{ \sigma _r}]_{L} $ & $[\frac{\sigma _{n\ell }}{\sigma
_r}]_{M}$  & $[\frac{ \sigma _{n\ell }}{\sigma _r}]_{H} $ &
$[\frac{Z \sigma _{n\ell }}{\sigma _r}]_{L}$ & $[\frac{Z \sigma
_{n\ell }}{\sigma _r}]_{M}$ & $[\frac{ Z \sigma _{n\ell }}{\sigma
_r}]_{H}$ \\ \hline
 1s  & 34.56  & 0.0006  & 0.0041 & 0.0047 & 0.034   & 0.221  & 0.255  \\
 2s  & 5.45   & 0.0224  & 0.0271 & 0.0271 & 1.211  & 1.461  & 1.463  \\
 2p  & 4.89   & 0.0703  & 0.0834 & 0.0836 & 3.796  &  4.506 & 4.513  \\
 3p  & 0.96   & 0.1017  & 0.1050 & 0.1050 & 5.492  &  5.670 & 5.670  \\
 3d  & 0.68   & 0.1734  & 0.1775 & 0.1775 & 9.364  &  9.585 & 9.585  \\
\hline
\end{tabular}
\end{center}
\end{table}

 The number of the K shell (1s shell) holes per recoil increases as the WIMP mass increases.
 It is quite sizable and remains
almost constant for  mass $\ge$ 100 GeV. This is due to the fact
that the average nuclear recoil velocity increases with the WIMP
mass. The number of the L shell ( 2s and 2p) holes per recoil can
be as large as 5-7 for  WIMP mass in the range  $30-300$ GeV.
Since K X-ray emission necessarily requires larger energy transfer
from WIMPs than L X-ray one, ratios of K X-ray to L X-rays depend
on the WIMP mass.  Accordingly, the K to L ratio can be
used to get the WIMP mass, if both can  be clearly observed.

 The $n \ell $ X-ray production rate is simply obtained by using the X-ray
 branching ratio as
 \begin{equation}
 \frac {\sigma _{n\ell }(X)}{\sigma _r} = b_{n\ell } \frac {\sigma _{n\ell }}{\sigma _r},
 \end{equation}
 where $\sigma _{n\ell }(X)$ is the sum of the X ray rates for X-rays
 filling the $n \ell $ shell and  $b_{n\ell }$ is the fluorescence ratio.
 The Auger electron branching ratio is simply given by 1-$b_{n\ell }$.
 Here we assumed that the inner-shell electron holes are
 filled by outer-shell electrons in the same atom via X-ray
 emission or the Auger effect. Non-radiative electron transfer
 to the inner-shell from neighboring atoms is considered to be small,
 since the nuclear recoil velocity is much smaller than the K-shell electron velocity.
 The K shell fluorescence ratio for Xe is 0.89 in case of one K-hole in
 the atom. X-ray cross-sections relative to the nuclear
cross-sections, given by $[Z \sigma_{nl}(X)/ \sigma_r]$
 are 0.03, 0.20, and 0.23  for WIMPs with 30, 100 and 300 GeV, respectively.
 The X-ray rate increases as the WIMP mass increases as can be seen
 from the data of   Table \ref{t:1}.

  The above results were obtained
  \cite{mou05} using  realistic wave functions
  \cite{Bunge-93}. It should be mentioned, however, that these
  wave functions do not include relativistic effects, which may
  somewhat affect the inner shell electrons and, in particular, $1s$
  bound electron wave functions.
  In the present
  case, however, the outgoing electrons have very low energy. So
  the event rates are not affected by the lower component of the
  corresponding spinors. We find that, in the case of the Coulombic
  interaction, the modification of the upper component due to relativistic effects
reduces the $1s$  hole production by less than $30~\%$.
  Anyway more detailed such calculations are under study and will appear
  elsewhere.

 K X-rays are followed by L X-rays, and L X-rays are by M X-rays and so on.
 The energy sum of these X-rays is just the K shell binding energy.
 These X-rays are converted to electrons via the photo-electric effect
 in detectors.
 Then, in the case of one $n \ell $-shell hole, the sum of the photo-electron energies and the sum of the Auger
 electron energies are given by the binding energy (-$\epsilon _{n\ell } \ge 0$).
 Therefore, one expects to find  a $n\ell $
excitation signal with a relative rate (ratio) of $Z \sigma _{n\ell }/\sigma_r$
and electron energy of -$\epsilon _{n\ell }$.

 It is indeed impressive to find that the average number of inner shell electron holes,
 per nuclear recoil, are about 0.2 in the K-shell and
about 5 in the L shell in the case of $^{131}$Xe for a WIMP mass of $100$ GeV. In other words, one recoil nucleus
is followed approximately one K X-ray per $5$ recoils and  by five L X-rays per recoil. This
effect, however, is less dramatic in the case of a lighter WIMP (see Table \ref{t:1}).\\
The implications and the impact of the X-ray signal  for high sensitivity
studies of WIMPs are great because of its unique features. Let us, first, discuss
 effects of the X-rays on the energy spectrum in inclusive
experiments. The total energy signal for one event of an elastic
scattering of WIMP is given by the sum,
\begin{equation}
E = Q~E_r + E_X + E_e,
\end{equation}
where $E_r$ is the recoil energy with the quenching factor $Q$ and
$E_X$ and $E_e$ are, respectively, the total energy of the X-rays
and that of the ionization and Auger electrons. In inclusive
experiments, where X-ray signals are not separated from Auger
electron signals, the energy sum of the X-rays and Auger-electrons
is measured. A large fraction of the recoil events are followed by
one K X-ray and/or several lower-energy X-rays, and thus the total
energy is shifted by the sum of the X-ray energies.

In case of the 1s electron ionization followed by K X-rays, the
energy spectrum is given approximately as
\begin{equation}
E_{1s} = Q~E^0_r + E_{X_{1s}}(1 - Q),
\end{equation}
where $E_{X_{1s}}=-\epsilon _{1s}$ is the sum of the X-ray energies
and $E^0_r = E_r + E_{X_{1s}} $ is the recoil energy without 1s
inner shell excitation. Note that the recoil energy is used partly
to excite the 1s electron to the continuum. Then the energy
spectrum is shifted to higher energy side by $E_{X_{1s}}(1 - Q)$.
This shift shows up as a bump at the energy around $E_{X_{1s}}(1 -
Q)$ in medium-heavy mass detectors with $Q \ll $ 1. These features
of the energy spectrum may be used to identify the WIMP events to
improve the detection sensitivity.

Exclusive studies of the X-rays are very powerful for high
sensitivity experiments. K X-rays from Xe fly through Xe detectors
for about 100 mg/cm$^2$, while L X-rays for about 4 mg/cm$^2$,
before depositing their energies via photoelectric effects. Thus
separation of the X-ray signals from those of the nuclear recoil
and the ionization electrons can be made by using good
position-resolution detectors of the order of 1 mg/cm$^2$ (i.e.
1.5 mm in 1 atm Xe gas) for exclusive measurements. The K X-rays
from the I isotopes have a range of about 200 $\mu$m in NaI.

The K$_{ij}$ X-ray ratio is evaluated as
\begin{equation}
\frac {\sigma _K (K_{ij})}{\sigma _r} = \frac {\sigma
_{1s}}{\sigma _r}~b_{1s} B(K_{ij}), \label{Kij-ratio}
\end{equation}
where $B(K_{ij})$ is the K-ij X-ray branch \cite{sal74}.
The K X-ray rates are evaluated for the K shell holes given in the Table 1
by using the K-ij X-ray branch for one K-hole in the atom.

The  cross-sections for K X-rays relative to the nuclear recoil
are obtained from the ratio in Eq. (\ref{Kij-ratio}) by
multiplying the total electron number $Z$. They are shown for
$^{131}$Xe isotopes in Table \ref{t:2}.


\vspace{0.5 cm}

\begin{table}[h]
\begin{center}
\caption{K X-ray cross sections relative to the nuclear recoil,
rates and energies in WIMPs nuclear interactions with $^{131}$Xe.
$[Z \sigma _K /\sigma _r]_L, [Z \sigma _K /\sigma _r]_M$ and $[Z
\sigma _K /\sigma _r]_H$ are the ratios for light (30 GeV), medium
(100 GeV) and heavy (300 GeV) WIMPs.} \label{t:2} \vspace{0.5cm}
\begin{tabular}{|cccccc|}
\hline K X-ray & $E_K(K_{ij})$ keV  & $B_K(K_{ij})$  & $[\frac{Z
\sigma _K(K_{ij})}{\sigma _r}]_{L}$ &  $[\frac{Z \sigma
_K(K_{ij})}{\sigma _r} ]_{M} $ & $[\frac{Z \sigma
_K(K_{ij})}{\sigma _r}]_{H} $ \\ \hline
 K$_{\alpha 2}$  & 29.5 & 0.284 & 0.0086 & 0.0560 & 0.0645 \\
 K$_{\alpha 1}$ & 29.8  & 0.527 & 0.0160 & 0.1036 & 0.1196 \\
 K$_{\beta 1}$  & 33.6  & 0.154 & 0.0047 & 0.0303 & 0.0350 \\
 K$_{\beta 2}$  & 34.4  & 0.034 & 0.0010 & 0.0067 & 0.0077 \\

\hline
\end{tabular}
\end{center}
\end{table}

\vspace{0.5 cm} K$_{\alpha}$ and K$_{\beta}$ lines can be
separated experimentally by using good energy-resolution
detectors, but the sum of all K lines can be measured in modest
energy-resolution experiments.

One option of detectors for exclusive studies of the X-rays following
WIMPs scattering off nuclei is a TPC with Xe gas.
The trajectory analysis makes it possible to identify the
WIMP nuclear interaction point with the recoil and ionization electrons
and the X-ray interaction point with the photo-electron track.
A super-module of Xe gas ionization chambers for nuclear recoils
and plastic scintillation-fibers for K X-rays is an alternative way
for exclusive studies of X-rays and nuclear recoils.

Recently highly-segmented NaI scintillator array has been
developed at Tokushima group \cite{fus05}. It consists of 16
layers of thin NaI plates, each with 50 mm long 50 mm wide and 500
$\mu $m thick. Since the thickness of the NaI plate is of the
order of the range of K X-rays from I isotopes,
nuclear recoils are measured in one layer of NaI in coincidence
with the  K X-rays in an adjacent layer. One may expect the
similar  K X-ray rate from $^{127}$I with $Z$ = 53 as the rate
from $^{131}$Xe  with $Z$ = 54 given above.

The exclusive experiments are free of most backgrounds from
natural and cosmogenic radioactive impurities in detector
components, detector shields and experimental walls. Cosmogenic
muons are well rejected by veto counters against charged
particles. Then remaining backgrounds in the exclusive experiments
are due to cosmogenic neutrons scattered off target nuclei,
resulting in inner-shell electron excitations and X-ray emissions.

The event rate for ionization electrons relative to that for
elastic neutron-nucleus scattering has already been estimated in
our previous work \cite{eji93}. Fig. (8) of this reference gives
less than $0.05$ per Kg per y for all produced electrons. Thus the
background K X-rays is expected to be less than that and even
less than 0.01 by means of active shields such as plastic or
liquid scintillators, since, in contrast to the weakly interacting
WIMPs, the neutrons are strongly interacting particles.  BG rates
from fast neutrons at 4000 m w.e underground laboratories are less
than 10$^{-3}$ counts per kg per day, which are negligible in
comparison with the WIMP signal rate resulting from a  10$^{-7}$
pb p-cross-section. So this sort of background is not worrisome.

 In any case K and L
X-rays as well nuclear recoils resulting from WIMP interactions
exhibit a similar pattern with those produced with neutrons
capable of causing the same momentum transfer as WIMPs in the mass
range of $100-300$ GeV, i.e. neutrons with energy in the $10-20$
MeV range. Thus one can  exploit this fact  to study
experimentally the signals of interest by using neutrons as a
probe.

Background events due to Compton and photoelectric scattering of
low-energy photons may give the same topology of energy deposition
as WIMP events followed by K or L X-rays if the Compton-scattered
electrons are in the nuclear recoil energy-window of 2-20 keV and
the Compton scattered photons are in the X-rays energy window.
Such low-energy photons, however, are unlikely to be Compton
scattered. Simulation of typical U-Th impurities  of the orders of
0.1 m Bq per kg give 10$^{-3}$ per kg per day, which is
negligible.  We should also mention  that low energy
photons less than 100 keV in
medium-heavy atoms are predominantly due to the  photoelectric effect, while
Compton scattering is unlikely. On the other hand, medium and higher
energy photons may double-Compton scattered to give two electron signals quite like
the recoils and the X-rays do. Scattered photons, however, are finally captured
in other parts of the detector or by veto-counters around the detector, and, thus,
they can be rejected.

Acutely, simulation analyses for thin NaI modules give BG rates of
the order of 2 $\times 10^{-2}$ per day per kg in the low energy
region in case of the exclusive measurements. This is similar to
the upper limit of high-sensitivity CDMS experiment \cite{ake03}
and 2 orders of magnitude lower than the DAMA BG rate. Accordingly
the sensitivity of the order of 10$^{-7}$ pb proton cross-section
can be expected in exclusive measurements for the medium and heavy
WIMPs.

In short, the X-rays following WIMP nuclear interactions are of
great interest to improve the sensitivity of the dark matter
studies. WIMPs interacting with nuclei in medium and heavy mass
region are likely followed by energetic K and L X-rays in the 10
keV region far above threshold energy of most detectors.  In
case of 100 GeV WIMPs interacting with $^{131}$Xe, the K
X-ray probability is more than 20 $\%$,
and the energy sum is as large as 34 keV. Thus inclusive study of
the recoil energy spectrum by means of solid detectors with a
large quenching factor can be used to measure effectively the hard
X-rays without the quenching reduction.

 Exclusive studies of the hard X-rays in coincidence with
 the nuclear recoil and ionizing electrons are very powerful for WIMP search.
 The X-ray shows up as an isolated peak in the energy spectrum, and
 the coincidence measurement makes it possible to be almost free of BG's.
 K X-rays are quite promising to search for medium and heavy WIMPs, and
 L X-rays are used for light WIMPs as well as medium and heavy WIMPs.
 Thus it is quite realistic to study WIMPs/LSP in the 10$^{-7}$ pb p-cross-section.

It should be noted that the fluorescence ratio and the K X-ray branching ratio used in the above discussions are those for one K hole in the atom. Actually, the outer-shell electron configuration in the recoil nucleus is not simple, but is  rather complex. In fact the K X-rays are mainly the K$_{\alpha 1}$ from the L$_{3}$ shell and the K$_{\alpha 2}$ from the L$_{2}$ shell, and their branches depend on the electron occupation-probabilities in the L$_{3}$  and L$_{2}$ shells. Actually, the K$_{\alpha 1}$ and K$_{\alpha 2}$ energies are so close to each other that they are not separated in most experiments. Then the  sum of the K X-ray intensities is proportional to the K shell vacancy-probability, and is insensitive to the L$_{3}$  and L$_{2}$ shell occupation-probabilities. In practice, the K X-ray ratio can be calibrated experimentally by using nuclear recoils from low-energy neutron scattering off the target nuclei to be used for WIMPs experiments.
Measuring angles of the  scattered neutrons, one gets the nuclear recoils corresponding to WIMPs in the 30 - 300 GeV range.

Acknowledgments: The first author ( H. E) thanks the Institute for
Nuclear Theory at the University of Washington and the Department
of Energy for partial support of this work. The second author
(Ch.C. M.) acknowledges support by the Greek State Grants
Foundation (IKY) under contract (515/2005).
 Finally (J.D. V.) is indebted to the Greek Scholarship
Foundation (IKYDA) for partial support, the Humboldt foundation for the Research Award
and to Professor Faessler for his
hospitality in Tuebingen.


\begin{thebibliography}{99}
\bibitem{san02}
M.G.~Santos et al., Phys. Rev. Lett. {\bf 88} (2002) 241302, and
references therein.
\bibitem{mau02}
P.D.~Mauskopf et al., Astrophys. J. {\bf 536} (2002) L59.
\bibitem{hal02}
N.W. Halverson et al., Astrophys. J. {\bf 568} (2002) 38,
\bibitem{spe03}
D.N.~Spergel et al., Astrophys. J. Suppl. {\bf 148} (2003) 175.
\bibitem{teg04}
M. Tegmark et al., Phys. Rev. D {\bf 69} (2004) 103501.
\bibitem{ber98}
R. Bernabei et al., Phys. Lett. B {\bf 424} (1998) 195, {\bf 450}
(1999) 448.
\bibitem{ben02}
A~Benoit et al. EDELWEISS collaboration, Phys. lett. B {\bf 545}
(2002) 43, V.~Sanglar et al. EDELWEISS collaboration, arXiv:
astro-ph/0306233.
\bibitem{ake03}
D.S.~Akerb et al. CDMS collaboration, Phys. Rev. D {\bf 68} (2003)
082002, and arXiv: astro-ph/0405033.
\bibitem{ver03}
J.D.~Vergados, Phys. Rev. D {\bf 67} (2003) 103003; J. Phys. G:
Nucl. Part. Phys. {\bf 30} (2004) 1127.
\bibitem{ver04}
J.D. Vergados, J. Phys. G: Nucl. Part. Phys. {\bf 30} (2004) 1127;
hep-ph/0406134.
\bibitem{eji93}
H.~Ejiri, K.~Fushimi and H.~Ohsumi, Phys. Lett. B {\bf 317} (1993)
14.
\bibitem{ver05}
J.D.~Vergados, P.~Quentin and D.~Strottman, Int. J. Mod. Phys. E14 (2005) 751;
arXiv: hep-ph/0310365.
\bibitem{ver05a}
J.D.~Vergados and H.~Ejiri, Phys. Lett. B {\bf 606} (2005) 313;
arXiv: hep-ph/0401151.
\bibitem{mou05}
Ch. C.~Moustakidis, J.D.~Vergados and H.~Ejiri, Nuc. Phys. B {\bf 727} (2005) 406; arXiv: hep-ph/0507123.
\bibitem{Bunge-93} C.F. Bunge, J.A. Barrientos, and A.V. Bunge,
At. Data Nucl. Data Tables  53 (1993) 113.
\bibitem{sal74}
S.I.~Salem, S.L.~Panossian and R.A.~Krause, Atomic and Nuclear Data Table {\bf 14} (1974) 91.\\
W.~Bambynek et al., Rev. Mod. Phys. {\bf 44} (1972) 716.
\bibitem{fus05}
K. Fushimi et al, private communication 2005.

\end{thebibliography}
\end{document}